\documentclass[12pt]{JHEP3}
\usepackage{amsmath,epsfig,amsfonts}

\def\be{\begin{equation}}
\def\ee{\end{equation}}
\def\bea{\begin{eqnarray}}
\def\eea{\end{eqnarray}}

\def\yzero{\smash{\hbox{$y\kern-4pt\raise1pt\hbox{${}^\circ$}$}}}
\def\p{\partial}

\def\be{\begin{equation}}
\def\ee{\end{equation}}
\def\bea{\begin{eqnarray}}
\def\eea{\end{eqnarray}}

\def\-{\hphantom{-}}

\def\s2{\frac{1}{\sqrt2}}

\def\beq{\begin{equation}}
\def\eeq{\end{equation}}
\def\beqa{\begin{eqnarray}}
\def\eeqa{\end{eqnarray}}

\def\IF{\relax{\rm I\kern-.18em F}}
\def\II{\relax{\rm I\kern-.18em I}}
\def\IP{\relax{\rm I\kern-.18em P}}
\def\IC{\relax\hbox{\kern.25em$\inbar\kern-.3em{\rm C}$}}
\def\IR{\relax{\rm I\kern-.18em R}}

\def\Dsl{\,\raise.15ex\hbox{/}\mkern-13.5mu D} 
\def\IZ{Z\kern-.4em  Z}

\title{A $\mathcal{N}=8$ action for multiple M2-branes with an arbitrary number of colors}

\author{M. P. Garc\'\i a del Moral $^1$, A. Restuccia $^2$\,\footnote{E-mail:
\emph{garciamormaria@uniovi.es; arestu@usb.ve}}\\
$^1$ Dipartimento di Fisica Teorica, Universit\`a di Torino \\
and INFN - Sezione di Torino; Via P. Giuria 1; I-10125 Torino,
Italy.\\ $\&$
Departamento de F\'{\i}sica, Universidad de Oviedo;\\
c. Calvo Sotelo 18, 33007, Oviedo, Spain. \\
$^2$
Departamento de F\'\i sica, Universidad Sim\'on Bol\'\i var\\
Apartado 89000, Caracas 1080-A, Venezuela}

\abstract{We obtain a $U(M)$ action for supermembranes
 with central charges in the Light Cone Gauge (LCG).
The theory
realizes all of the symmetries and constraints of the supermembrane
together with the invariance under a U(M)
gauge group with $M$ arbitrary. The worldvolume action has
 (LCG) $\mathcal{N}=8$ supersymmetry and it corresponds to $M$ parallel
supermembranes minimally immersed on the target $M_9$x$T^2$ (MIM2).
In order to ensure the invariance under the symmetries and to close the
corresponding
algebra,  a star-product  determined by the central charge condition is introduced.
It is constructed with a nonconstant symplectic two-form where curvature terms
 are also present. The theory is in the strongly coupled gauge-gravity regime.
At low energies, the theory enters in a decoupling limit and it is described by an
ordinary $\mathcal{N}=8$ SYM in the IR phase for any number of M2-branes.
We analyze also other limits of the
theory:  We consider the number of supermembranes going to infinity.
All of the symmetries are realized in this case with
the ordinary product and they correspond to a condensate of
M2-branes.  If we consider, instead of the exact action, its matrix
regularized version, for suitable large $N$ but with a finite number of
colors, the algebra of the first class-constraints also closes with the ordinary product and we
obtain a regularized 2+1D $SU(M)$ SYM theory  in the IR phase that
is coupled to the $SU(N)$ regularized supermembrane action. The
supersymmetric regularized theory with the ordinary product has purely discrete spectrum.
 Our theory with the
star-product is a full-fledged description of multiple M2-branes minimally
immersed in distinction with other constructions that represent
multiple M2-branes in the low energy approximation. }

 \preprint{FFUOV-09/02;DFTT-33/2009}


\begin{document}

\section{Introduction}

The supermembrane with a topological restriction associated to
an irreducible winding has been shown to have very interesting
properties: discreteness of the supersymmetric spectrum \cite{Boulton:2002br, Boulton:2004yt,Boulton:2006mm}, spontaneous
breaking of supersymmetry, stabilization of most of the moduli \cite{GarciadelMoral:2007xj}, a spectrum containing dyonic
strings plus pure supermembrane excitations \cite{GarciadelMoral:2008qe},
formulation on a G2 manifold \cite{Belhaj:2008qz}.
This restriction can be seen at algebraic level as  a central charge
condition on the 11D  supersymmetric algebra and geometrically as a
condition of being minimally immersed into the target space \cite{Bellorin:2005kj}, so from
now on, we will denote it as MIM2. The goal of this paper is to
consistently extend  this action of a single MIM2 to a theory of
interacting parallel M2-branes minimally immersed (MIM2's) preserving all of the symmetries of the theory:
supersymmetry and invariance under area preserving
 diffeomorphisms. The theory is not conformal invariant. In the extension the
  gauge and gravity sectors are strongly correlated. It corresponds
to have a M-theory dual of the Non-Abelian Born-Infeld action
describing a bundle of multiple D2-D0 branes, so  we work in the high energy
approximation. When the energy scale is low, the theory decouples and
it is effectively described by a $\mathcal{N}=8$ SYM in the IR phase.
As the energy scale raises the YM coupling constant becomes weaker and at some point,
oscillations modes of the pure supermembrane appear and the theory enters in the strong correlated gauge-gravity sector.\newline

Recently there has been a impressive
amount  of work trying to extend an effective action for the
multiple M2-branes. The original motivation was to prove Maldacena's
Conjecture for M-theory \cite{Maldacena:1997re} according to which
M-theory/$AdS_4\times S^7$ should be dual to a $CFT_3$ generated by
the action of multiple M2-branes in the decoupling limit, that is ,
for a large number $M$ of M2's. This action was postulated
as a low energy description of a SU(M) formulation of the
multiple action of multiple M2-branes.
The low energy effective action of the multiple M2-branes was
expected to correspond to the IR limit of a SYM theory defined in 3D
since M-theory is strongly coupled. Contrary to the IIB AdS-CFT
duality, for the 11D supermembrane it was necessary to  introduce
non-dynamical gauge fields via a Chern Simons term to avoid breaking
the matching of the bosonic and fermionic degrees of freedom. The
first attempt was due to Schwarz who developed $N=1$ and $N=2$ cases
but could not find a N=8 susy action for supermembranes \cite{Schwarz:2004yj}.
The symmetries imposed on the action were the $\mathcal{N}=8$
worldvolume superconformal action and SU(N) symmetry.
\cite{Bagger:2006sk},\cite{Bagger:2007jr} and independently \cite{Gustavsson:2007vu} were the
first to obtain a realization of this algebra by imposing fields
to be evaluated on a three algebra with positive inner metric, (a particular case of Fillipov
algebras)\footnote{Interesting works on these algebras in relation with the supermembrane theory were obtained
long time ago in \cite{Hoppe:1996xp}, and it was applied to the M5 in \cite{DeCastro:2003jc}. In those case instead of Fillipov
 algebras Nambu-Poisson algebras are needed and there are important subtleties concerning its quantization in odd dimensions.
 See for example\cite{Curtright:2002fd,Curtright:2003wj} and \cite{Kawamura:2002yz} for a cubic matrix description.}.
 This three-algebra can only be
realized in terms of a unique finite dimensional gauge group $SO(4)$
for an inner positive metric, \cite{Ho:2008bn} with a twisted
Chern-Simons terms, see also \cite{Bonelli:2008kh}. This 3-algebra can also be re-expressed as the
tensor product of two $SU(2)\times SU(2)$ gauge groups associated
each one to a different Chern-Simons term \cite{VanRaamsdonk:2008ft}. Consistency checks of BLG in the funnel were indicated in \cite{Krishnan:2008zm} as well as other properties of Lorentzian 3-algebras. To
realize this duality in the decoupling limit it is necessary to
obtain a large number of supermembranes for an arbitrary number of
M2-branes.In order to improve this situation and generalize it for
general $SU(M)$ gauge groups, different avenues have been followed:
The most sucesfull has been to formulate  a Chern-Simons-matter
theory $\mathcal{N}=6$ by \cite{Aharony:2008ug} in which they are able to
generalize the theory to an arbitrary $SU(N)$ and recover also BLG
theory for the case of $N=2$. The ABJM, or at least a sector of it,
can be also recover from the 3-algebra formulation by relaxing the
condition of total antisymmetry of the structure constants
\cite{Bagger:2008se} In a serie of
papers it has also been explored the possibility of obtaining
$\mathcal{N}=8$ models with arbitrary gauge groups  by relaxing the
positivity condition of the internal metric \cite{Gomis:2008uv},\cite{Bandres:2008vf},\cite{Gomis:2008be}.
These models have successfully obtained generic SU(N) gauge groups
but to the price of lack of unitarity because of the presence of
ghosts. Ghost-free
actions have been formulated \cite{Bandres:2008kj} and has been shown to
exactly correspond to a reformulation of a $N=8$ Super Yang Mills in
$(2+1)$D and not to its IR limit \cite{Ezhuthachan:2008ch}. The relation between multiple M2-branes and D2-branes has been also analyzed in,
\cite{Mukhi:2008ux,Ho:2008ei,Pang:2008hw,Li:2008ya}.
A non-linear realization of Lorentzian algebras has been proposed in \cite{Iengo:2008cq}.
Massive deformations as for example,
\cite{Song:2008bi,Hyun:2003se,Hosomichi:2008qk} or \cite{Lee:2008cr}
for topological twisting have also been considered.
These superconformal models can also be obtained
by taking the conformal limit  of gauge supergravities in 3D
\cite{Bergshoeff:2008bh}. This seems to indicate that the information of all of these theories
could be contained on supergravity\footnote{We thank H. Nicolai for comments to this respect.}.
 This is in some sense surprising
for a description that is intended to describe the quantum
formulation of multiple M2-branes or even the infrared physics of a
Yang-Mills theory. There has been recent advances focused
in models with less number of supersymmetries, for example, \cite{Franco:2008um}.\newline
We would like to emphasize that in what follows,
we consider, a non abelian extension of the full-fledged theory
 describing multiple M2-branes in the L.C.G. without imnposing the conformal symmetry not present in the original theory of the M2-brane.  \newline

The paper is organized in the following way: In Section 2 we make a
short summary of the formulation and main properties of the
supermembrane minimally immersed. On section 3 we introduce a non
abelian extension of the MIM2 that allows to consider different
limits, a matrix model regularization with a finite and arbitrary
number of colors as well as a condensate of M2's in the large N
matricial limit. In order to have a nonabelian formulation in 2+1D
respecting all of the symmetries (in particular invariance under
diffeomorphisms preserving the area) of the former theory for an
arbitrary number of colors a noncommutative star-product has to be
included to close the algebra. This is explained in Section 4. This
noncommutative star product differs from the Seiberg-Witten map
since the noncommutative parameter is non constant on the spatial
variables. In section 5 we analize its supersymmetry and we show
that it has $\mathcal{N}=8$ supersymmetries.  To conclude we finally present our
results and main properties, emphasizing its potential phenomenological
interest.
\section{D=11 Supermembrane with  central charges on a $M_{9}\times T^{2}$ target manifold}
In this section we will make a self-contained summary of the construction of the minimally immersed M2-brane (MIM2).
The hamiltonian of the $D=11$ Supermembrane \cite{Bergshoeff:1987cm} may be
defined in terms of maps $X^{M}$, $M=0,\dots, 10$, from a base
manifold $R\times \Sigma$, where $\Sigma$ is a Riemann surface of
genus $g$ onto a target manifold which we will assume to be 11D
Minkowski. The canonical reduced hamiltonian
to the light-cone gauge has the expression \cite{deWit:1988ig}

\begin{equation}\label{equ1}
  \mathcal{H}= \int_\Sigma  d\sigma^{2}\sqrt{W} \left(\frac{1}{2}
\left(\frac{P_M}{\sqrt{W}}\right)^2 +\frac{1}{4} \{X^M,X^N\}^2+
\overline{\Psi}\Gamma_{-}\Gamma_{M}\{X^M,\Psi\}\right)
\end{equation}
subject to the constraints \begin{equation}  \label{e2}
\phi_{1}:=d(\frac{P_{M}}{\sqrt{W}}dX^{M} +\overline{\Psi}\Gamma_{-}d\Psi)=0 \end{equation} and
\begin{equation} \label{e3}
\phi_{2}:=
   \oint_{C_{s}}(\frac{P_M}{\sqrt{W}}dX^M +\overline{\Psi}\Gamma_{-}d\Psi)= 0,
\end{equation}
where the range of $M$ is now $M=1,\dots,9$ corresponding to the
transverse coordinates in the light-cone gauge, $C_{s}$,
$s=1,\dots,2g$ is a basis of  1-dimensional homology on $\Sigma$,
 \be \label{e4}\{X^{M}, X^{N}\}= \frac{\epsilon
^{ab}}{\sqrt{W(\sigma)}}\partial_{a}X^{M}\partial_{b}X^{N}. \ee
$a,b=1,2$ and $\sigma^{a}$ are local coordinates over $\Sigma$.
$W(\sigma)$
 is a scalar density introduced in the light-cone gauge fixing procedure.
 $\phi_{1}$ and $\phi_{2}$ are generators of area preserving diffeomorphisms, see \cite{deWit:1989vb}. That is
\be \sigma\to\sigma^{'}\quad\to\quad W^{'}(\sigma)=
W(\sigma).\nonumber \ee When the target manifold is simply connected
$dX^{M}$ are exact one-forms.

The spectral properties of (\ref{equ1}) were obtained in the context of a
 $SU(N)$ regularized model \cite{deWit:1988ig} and it was shown to
have continuous spectrum from $[0,\infty)$.

This property of the theory relies on two basic facts: supersymmetry
and the presence of classical singular configurations, string-like
spikes, which may appear or disappear without changing the energy of
the model but may change the topology of the world-volume. Under
compactification of the target manifold generically the same basic
properties are also present and consequently the spectrum should be
also continuous \cite{deWit:1997zq}. In what follows we will impose a
topological restriction on the configuration space. It characterizes
a $D=11$ supermembrane with non-trivial central charges generated by
the wrapping on the compact sector of the target space
\cite{GarciadelMoral:2001zb},\cite{Boulton:2001gz},\cite{Boulton:2002br},\cite{Boulton:2006mm}.
We will consider in this paper the case $g=1$ Riemann
surface as a base manifold $\Sigma$ on a $M_9$x$T^2$ target space.
The configuration maps satisfy:

\begin{equation}\label{e5}
 \oint_{c_{s}}dX^{r}=2\pi L_{s}^{r}R^{r}\quad r,s=1,2.
\end{equation}\begin{equation}\label{e7}
 \oint_{c_{s}}dX^{m}=0 \quad m=3,\dots,9
\end{equation}
\\

 where $L^{r}_{s}$ are integers and $R^{r}, r=1,2$ are the radius of $T^{2}$.
  This conditions ensure that we are mapping $\Sigma$ onto a
$T^2$ sector of the target manifold.

We now impose the central charge condition \be\label{e8}
I^{rs}\equiv \int_{\Sigma}dX^{r}\wedge dX^{s}=(2\pi
R_{1}R_{2})n\epsilon^{rs} \ee where $\omega^{rs}$ is a symplectic
matrix on the $T^{2}$ sector of the target and $n=det L_i^r$ represents the irreducible winding.

 The topological condition
(\ref{e8}) does not change the field equations of the hamiltonian
(\ref{equ1}).
  In fact, any variation of $I^{rs}$ under a change $\delta X^{r}$, single valued over
 $\Sigma$, is identically zero. In addition to the field equations obtained from (\ref{equ1}),
the classical configurations must satisfy the condition (\ref{e8}).
It is only a topological restriction on the original set of
classical solutions of the field equations. In the quantum theory
the space of physical configurations is also restricted by the
condition (\ref{e8}). The geometrical interpretation of this
condition has been discussed in previous work
\cite{Martin:1997cb},\cite{Martin:2001te}. We noticed that (\ref{e8}) only
restricts the values of
 $L_{s}^{r}$, which are already integral numbers from (\ref{e5}).\\

We consider now the most general map satisfying condition
(\ref{e8}). A closed one-forms $dX^{r}$ may be decomposed into the
harmonic plus exact parts: \begin{equation}
dX^{r}=M_{s}^{r}d\widehat{X}^{s}+dA^{r}
\end{equation}where $d\widehat{X}^{s}$, $s=1,2$ is a basis of
harmonic one-forms over $\Sigma$ and $dA^{r}$ are exact
one-forms. We may normalize it by choosing a
canonical basis of homology and imposing

\begin{equation} \oint_{c_{s}}d\widehat{X}^{r}=\delta_{s}^{r}. \end{equation} We have now
considered a Riemann surface with a class of equivalent canonical
basis. Condition (\ref{e5}) determines \begin{equation}\label{eq1}
M_{s}^{r}=2\pi R^{r}L_{s}^{r},\end{equation}  we rewrite $L^r_s=l_r S_s^r$ and $l_1.l_2=n$.
We now impose the condition (\ref{e8}) and obtain
\begin{equation} S_{t}^{r}\omega^{tu}S_{u}^{s}=\omega^{rs}, \end{equation} that
is, $S\in
Sp(2,Z)$. This is the most general map satisfying (\ref{e8}).
See \cite{GarciadelMoral:2008qe} for details, in particular for $n>1$.\\

 The
natural choice for $\sqrt{W(\sigma)}$ in this geometrical setting
is to consider it as the density obtained from the pull-back of the
Kh\"aler two-form on $T^{2}$. We then define \begin{equation}
\sqrt{W(\sigma)}=\frac{1}{2}\partial_{a}\widehat{X}^{r}\partial_{b}\widehat{X}^{s}\omega_{rs}.
\end{equation}

$\sqrt{W(\sigma)}$ is then invariant under the change
\begin{equation} d\widehat{X}^{r}\to S_{s}^{r}d\widehat{X}^{s}, \quad
S\in Sp(2,Z) \end{equation}

But this is just the change on the canonical basis of harmonics
one-forms when a biholomorphic map in $\Sigma$ is performed changing
the canonical basis of homology. That is, the biholomorphic (and
hence diffeomorphic) map associated to the modular transformation on
a Teichm\"uller space. We thus conclude that the theory  is
invariant not only under the diffeomorphisms generated by $\phi_{1}$
and $\phi_{2}$, homotopic to the identity, but also under the diffeomorphisms, biholomorphic
maps,
changing the canonical basis of homology by a modular transformation.\\
Having identified the modular invariance of the theory we may go
back to the general expression of $dX^{r}$, we may always consider a
canonical basis such that \begin{equation}
dX^{r}=2\pi l^rR^{r}d\widehat{X^{r}}+dA^{r}. \end{equation} the
corresponding degrees of freedom are described exactly by the
single-valued
 fields $A^{r}$. After replacing this expression in the hamiltonian (\ref{equ1}) we obtain,

\begin{equation}\label{e9}
 \begin{aligned}
H&=\int_{\Sigma}\sqrt{W}d\sigma^{1}\wedge
d\sigma^{2}[\frac{1}{2}(\frac{P_{m}}{\sqrt{W}})^{2}+\frac{1}{2}
(\frac{\Pi^{r}}{\sqrt{W}})^{2}+
\frac{1}{4}\{X^{m},X^{n}\}^{2}+\frac{1}{2}(\mathcal{D}_{r}X^{m})^{2}+\frac{1}{4}(\mathcal{F}_{rs})^{2}
\\&+(n^2 \textrm{Area}_{T^2}^2)+ \int_{\Sigma}\sqrt{W}\Lambda
(\mathcal{D}_{r}(\frac{\Pi_{r}}{\sqrt{W}})+\{X^{m},\frac{P_{m}}{\sqrt{W}}\})]\\& + \int_{\Sigma}
\sqrt{W} [- \overline{\Psi}\Gamma_{-} \Gamma_{r}
\mathcal{D}_{r}\Psi -\overline{\Psi}\Gamma_{-}
\Gamma_{m}\{X^{m},\Psi\}-
 \Lambda \{ \overline{\Psi}\Gamma_{-},\Psi\}]
 \end{aligned}
 \end{equation}
where  $\mathcal{D}_r X^{m}=D_{r}X^{m} +\{A_{r},X^{m}\}$,
$\mathcal{F}_{rs}=D_{r}A_s-D_{s }A_r+ \{A_r,A_s\}$, \\
 $D_{r}=2\pi l_r
R_{r}\frac{\epsilon^{ab}}{\sqrt{W}}\partial_{a}\widehat{X}^{r}\partial_{b}$
and $P_{m}$ and $\Pi_{r}$ are the conjugate momenta to $X^{m}$ and
$A_{r}$ respectively. $\mathcal{D}_{r}$ and $\mathcal{F}_{rs}$ are
the covariant derivative and curvature of a symplectic
noncommutative theory \cite{Martin:1997cb},\cite{Boulton:2001gz}, constructed from
the symplectic structure $\frac{\epsilon^{ab}}{\sqrt{W}}$ introduced
by the central charge. The last term represents its
supersymmetric extension in terms of Majorana spinors. The physical
degrees of the theory are the $X^{m}, A_{r}, \Psi_{\alpha}$ they are
single valued fields on $\Sigma$.

\subsection{Quantum supersymmetric analysis of a single MIM2}

We are going to summarize the spectral properties of the above
hamiltonian.  The bosonic potential of the (\ref{e9}) satisfies the
following inequality \cite{Boulton:2006mm} ( in a particular gauge
condition)
\begin{equation}\label{u3}
 \begin{aligned}
&\int_{\Sigma}\sqrt{W}d\sigma^{1}\wedge d\sigma^{2}[
\frac{1}{4}\{X^{m},X^{n}\}^{2}+\frac{1}{2}(\mathcal{D}_{r}X^{m})^{2}
+\frac{1}{4}(\mathcal{F}_{rs})^{2}]\\ \nonumber & \ge
\int_{\Sigma}\sqrt{W}d\sigma^{1}\wedge
d\sigma^{2}[\frac{1}{2}(\mathcal{D}_{r}X^{m})^{2}+(\mathcal{D}_{r}A_s)^2]
 \end{aligned}
 \end{equation}
 The right hand member under regularization
 describes a harmonic oscillator potential. In particular,
 any finite dimensional truncation of the original infinite
  dimensional theory satisfies the above inequality. We consider regularizations satisfying the above inequality.
 We denote  the regularized hamiltonian of the supermembrane with
the topological restriction by $H$, its bosonic  part $H_b$ and its
fermionic potential $V_f$, then
\begin{equation} H = H_b+V_f.\end{equation}
We can define rigorously the domain of $H_b$ by means of Friederichs
extension techniques. In this domain $H_b$ is self adjoint and it
has a complete set of eigenfunctions  with eigenvalues accumulating
at infinity. The operator multiplication by $V_f$ is relatively
bounded with respect to $H_b$. Consequently  using Kato perturbation
theory  it can be shown that $H$ is self-adjoint if we choose
\begin{equation}
Dom{H}=Dom{H_b}.\end{equation}
In \cite{Boulton:2002br} it was shown that H possesses a complete set of
eigenfunctions and its spectrum is discrete, with finite
multiplicity and with only an accumulation point at infinity. An
independent proof was obtained in \cite{Boulton:2004yt} using the spectral
theorem and theorem 2 of that paper. In section 5 of \cite{Boulton:2004yt} a
rigorous proof of the Feynman formula  for the Hamiltonian of the
 supermembrane was obtained.
In distinction, the hamiltonian of the supermembrane, without the
topological restriction, although it is positive, its fermionic
potential is not bounded from below and it is not a relative
perturbation of the bosonic hamiltonian. The use of the Lie product
theorem in order to obtain the Feynman path integral is then not
justified. It is not known and completely unclear whether a Feynman
path integral formula exists for this case. In
\cite{Boulton:2006mm}  it was proved that the theory of the supermembrane with
central charges, corresponds to a nonperturbative
 quantization of a symplectic Super Yang-Mills in a confined phase and the theory
 possesses a mass gap.

 In \cite{Belhaj:2008qz}we constructed of the
supermembrane with the topological restriction on an orbifold with
$G_2$ structure that can be ultimately deformed to lead to a true G2
manifold. All the discussion of the symmetries on the Hamiltonian
was performed directly in the Feynman path integral, at the quantum
level, then valid by virtue of our previous proofs.

\section{A non-abelian extension of the MIM2-brane: A first attempt.}
In this section we show a first attempt to obtain a non abelian extension of the MIM2-brane.
It  requires to obtain a $U(M)$ or $SU(M)$ formulation of the MIM2 theory, for an arbitrary number of colors $M$.
We will see that naive extensions are unable to achieve it. Along this section we will characterize the compatibility
problem between the non abelian gauge group and the infinite group of diffeomorphisms preserving the area. This will
give us a better understanding on how this problem can be overcome, as it is shown in section 4. where a truly non abelian
extension can be found. The cases contained in this section correspond to particular limits of the general construction of Section 4.
Let us introduce first some preliminary definitions that will become of utility along the discussion.
\newline
We will denote $T_A$, $A=(a_1,a_2)$,
$a_1,a_2=-(N-1),\dots,(N-1)$, $(m,n)\ne (0,0)$, the generators of the
Weyl-Heisenberg group. They satisfy
\bea T_A ^{\dag}=T_{-A}\\
tr T_A=0\\
T_A T_B =N e^{\frac{i\pi(B\wedge A)}{N}T_{A+B}}\\
T_{(a_1+N,a_2)}=e^{i\pi a_2} T_{(a_1,a_2)}\\
T_{(a_1,a_2+N)}=e^{i\pi a_1}T_{(a_1,a_2)}\eea The algebra $su(N)$ may be
realized in terms of $T_A$. The generators of $su(N)$ may be
expressed as $i(T_A+T_A^{\dag}),(T_A-T^{\dag}_A)$. A real scalar
field $X$ with values on $su(N)$ may be expanded as \bea X=X^A T_A =
\frac{-i}{2}(X^A + \overline{X}^A)i(T_A+T_A^{\dag})+ \frac{1}{2}(X^A
-\overline{X}^A)(T_A-T_A^{\dag}), \eea where $\overline{X}^A$ is the
complex conjugate of $X^A$.

The generators of $su{(NM)}$ may be realized in terms of $T_A\otimes
H_b$, $T_A\otimes \mathbb{I}_{M}$, $\mathbb{I}_{N}\otimes H_b$,
where $T$ and $H$ are the Weyl-Heisenberg generators associated to
$su(N)$ and $su(M)$ respectively. We associate to each member of the
above basis the kronecker product of the corresponding matrices.
That is, to $H_a\otimes H_b$ the krocnecker product of the matrices
$T_A$ and $H_b$. The bracket of
 the elements of the basis is the corresponding anticommutator of the matrices. With these definitions $T_A\otimes H_b$, $T_A\otimes
\mathbb{I}_{M}$, $\mathbb{I}_{N}\otimes H_b$ are the generators of $su(NM)$. $T_A \otimes \mathbb{I}_{M}$
and $\mathbb{I}_{N}\otimes H_b$ are the generators of the algebra of the direct product group $SU(N)\times SU(M)$, a subalgebra of $su(NM)$.\newline

We have,\bea\begin{aligned}
&[T_A,T_B]=f_{AB}^{C}T_{C}, \quad
f_{AB}^{C}=2iNsen(\frac{(B\wedge A)\pi}{N})\delta_{A+B}^{C}\\
\nonumber &\{T_A,T_B\}=d_{AB}^{C}T_{C},\quad d_{AB}^{C}= 2N
cos(\frac{(B\wedge A)\pi}{N})\delta_{A+B}^{C} \end{aligned}\eea and
\bea \label{con1} [T_A\otimes H_a,T_B\otimes
H_b]=(f_{AB}^{C}d_{ab}^{c}+d_{AB}^{C}f_{ab}^{c})T_{C}\otimes H_c
\eea We can extend the range of the index $A$ or $a$, but not both
together, to include $(0,0)$.The corresponding matrix is then
defined as usual \bea T_{(0,0)}=N\mathbb{I}_{N}\quad or \quad
H_{(0,0)}=M\mathbb{I}_{M}.\eea The commutation relation (\ref{con1}) is
then valid for all generators $T_A\otimes H_b$, $T_A\otimes
H_{(0,0)}$, $T_{(0,0)}\otimes H_a$.

We will denote \bea\label{u1}
\mathbf{F}_{\mathbb{A}\mathbb{B}}^{\mathbb{C}}\equiv
f_{AB}^{C}d_{ab}^{c}+d_{AB}^{C}f_{ab}^{c} \eea where
$\mathbb{A}=(A,a),\mathbb{B}=(B,b),\mathbb{C}=(C,c)$.
$\mathbf{F}_{\mathbb{A}\mathbb{B}}^{\mathbb{C}}$ is totally
antisymmetric. It satisfies the Jacobi Identity and has the expression
\bea \mathbf{F}_{\mathbb{A}\mathbb{B}}^{\mathbb{C}}=4iMN
sin(\frac{(B\wedge A)\pi}{N}+\frac{(b\wedge
a)\pi}{M})\delta_{\mathbb{A}\mathbb{B}}^{\mathbb{C}}. \eea In the following, we intend
to extend the hamiltonian of the supermembrane to include $su(M)$
valued fields, preserving the number of physical degrees of freedom
(times the dimension of the internal algebra). The main point is to
extend the area preserving constraint leaving invariant its first
class property. The algebraic structure of the supermembrane
hamiltonian is provided by the symplectic bracket, \bea \{ X^m,
P_m\}=\frac{\epsilon^{ab}}{\sqrt{W}}\partial_a X^m \partial_b P_m.
\eea We consider an extension of it of the form \bea\label{con2}
\{X^{am},P_M^{b}\}d_{ab}^{c}+f_{ab}^{c}X^{ma}P_{m}^{b} \eea where
$f_{ab}^c$ and $d_{ab}^c$ are respectively the structure constant
tensor and the totally symmetric tensor of $su(M)$.

We will perform the analysis on a matrix regularized model. If we
now expand
 the scalars on the base manifold $\Sigma$ in terms of an
 orthonormal basis $Y_A$,
 \bea
X^{ma}(\sigma^1,\sigma^2,\tau)=\sum_{A=-\infty}^{+\infty}X^{maA}(\tau)Y_A (\sigma^1,\sigma^2)
 \eea

 and define as usual
 \bea
\{Y_A,Y_B\}=g_{AB}^{C} Y_C\quad Y_AY_B=\widetilde{d}_{AB}^{C}Y_C
 \eea
 where $g_{AB}^{C}$ is the structure constant of the algebra of APD
 and  $\widetilde{d}_{ABC}$ is the totally symmetric tensor of the APD
 algebra. We have just re-written the theory in its matrix form,
 integrating out the spatial dependence captured on the APD structure constants as usual \cite{deWit:1988ig},
 but without regularizing it at this stage.
 We obtain for (\ref{con2})
 \bea\
\widetilde{\mathbf{F}}_{\mathbb{A}\mathbb{B}}^{\mathbb{C}}=g_{AB}^{C}d_{ab}^c
+\widetilde{d}_{AB}^{C}f_{ab}^c .\eea
 The basis $Y_A$, for a
compact torus $\Sigma$, may be expressed in terms of the harmonic
functions $\widehat{X}^r, r=1,2$ of section 2, normalized by
$\int_{\mathcal{C}_s} d\widehat{X}^r=2\pi \delta_s^r$, as \bea
Y_{(a_1,a_2)}=e^{i(a_1 \widehat{X}^1+a_2 \widehat{X}^2)} \eea we
then have \bea \frac{1}{Vol(\Sigma)}\int_{\Sigma}
\sqrt{W}Y_{(a_1,a_2)}\overline{Y}_{(b_1,b_2)}=\delta_{(a_1,b_1)(a_2,b_2)},
 \eea
 and the APD tensors of the torus are,\bea\begin{aligned}
&g_{AB}^C=(B\wedge
A)(\frac{1}{2}\epsilon_{rs}\epsilon^{ab}\partial_a
\widehat{X}^r\partial_b\widehat{X}^s)\delta_{A+B}^{C}=(B\wedge
A)\delta_{A+B}^C.\\ \nonumber
&\widetilde{d}_{AB}^C=\delta_{A+B}^C.\end{aligned} \eea

We can regularize the model by truncating the infinite expansion and
allowing the fields to be valued in the adjoint representation of a
$SU(N)$ group, \cite{deWit:1988ig}. In the supermembrane theory with
a compactified sector of the target space it is not possible to
extend directly the regularization to the harmonic sector fields
\cite{deWit:1997zq}. However in the minimally immersed sector of the
compactified supermembrane, a well defined theory by itself, the
harmonic sector is completely determined and there exists a
consistent regularization of the theory
 \cite{GarciadelMoral:2001zb}. The harmonic sector is related to a global symmetry $SL(2,Z)$,
 realized as a diffeomorphisms not connected to the identity in the infinite dimensional theory and to
 the center of $SU(N)$ in the regularized case.

The regularized structure constants for the $SU(N)$ matrix model
are the standard ones \cite{deWit:1988ig,deWit:1989vb} in terms of $T_A$ generators with $A=1,\dots, N^2-1$.
In order to guarantee the appropriate convergence to the original
APD structure constants we re-scale $d_{ABC}$ by a factor of $\frac{1}{N}$,
and although we will do it, it is not necessary to impose this requirement to the color group since the color
index is not the regularized version of a theory in the continuum.

 If $B\wedge
A$ remains bounded
  and $N\to\infty$ we have \bea
lim_{N\to\infty}\frac{1}{2i} f_{AB}^C=g_{AB}^C,\quad
lim_{N\to\infty} \frac{1}{2N}d_{AB}^{C}=\widetilde{d}_{AB}^C.\eea

This limit was considered in
\cite{deWit:1988ig},\cite{deWit:1989vb},\cite{deWit:1997zq}. The
semiclassical supermembrane subject to an irreducible wrapping was
first analyzed in \cite{Duff:1987cs}.   The large $N$ limit of the
spectrum of the regularized $SU(N)$ model of the semiclassical
minimally immersed supermembrane was studied in
\cite{Boulton:2006mm}, and it was proven that its eigenvalues
$\lambda_N<E$  for a fixed energy $E$ converge to the eigenvalues
$\lambda<E$ of the semiclassical supermembrane theory, when $N\to
\infty$. The boundness condition of $B\wedge A$ is ensured by the
condition that only modes with energy less than $E$ are considered.
The large $N$ limit is taken with $E$ fixed. We consider now a
regularized model with gauge group $SU(NM)\supset SU(N)\times
SU(M)$, with algebraic structure represented as in  (\ref{u1}), that
is \bea [X,P]=X^{\mathbb{A}}P^{\mathbb{B}}
\mathbf{F}_{\mathbb{A}\mathbb{B}}^{\mathbb{C}}.\eea

The first remark  is that the first class constraint of the
supermembrane theory becomes a first class constraint of the gauge
theory. The algebra of the first class constraint is exactly the
algebra of the Gauss constraint of a (0+1)Yang-Mills theory with
gauge group $SU(NM)$.

The constraint becomes \bea
\phi^{\mathbb{B}}=\lambda_{r\mathbb{A}}^{\mathbb{B}}\Pi^{r
\mathbb{A}}+A_{r}^{\mathbb{A}}\Pi^{r\mathbb{C}}\mathbf{F}_{\mathbb{A}\mathbb{C}}^{\mathbb{B}}+
X^{m\mathbb{A}}P^{\mathbb{C}}_{m}\mathbf{F}_{\mathbb{A}\mathbb{C}}^{\mathbb{B}}
+\overline{\Psi}^{\mathbb{A}}\Gamma_{-}\Psi^{\mathbb{C}}\mathbf{F}_{\mathbb{A}\mathbb{C}}^{\mathbb{B}}=0
\eea
 $\lambda_{r\mathbb{A}}^{\mathbb{B}}$ is the truncated version of
 the corresponding APD tensor defined from $D_{r}Y_A$, since it is
 a scalar on $\Sigma$. It may be decomposed in terms of the basis
 $Y_A$,
 \bea
D_r Y_A=\lambda_{rA}^{B}Y_B.\eea The algebra of the first class
constraint is, in terms of parameters $\epsilon$, $\lambda$, \bea
[<\epsilon,\phi>, <\lambda,\phi>]_{P.B}=<[\epsilon,\lambda]\phi>
\eea
where $< >$ denotes integration on $\Sigma$.
The constraint contains a linear term on the (0+1)D fields in a
similar way as in Yang-Mills theories. This property ensures the
elimination from the constraint the gauge degrees of freedom on
$\Pi^r$ and the corresponding one from $A_r$ by an admissible gauge
fixing condition. We notice that from a supermembrane on a compact
base manifold there is no way to fix an angle variable, that is to
identify coordinates on the base manifold with coordinates on the
target-space since, angle variables are harmonic on the base
manifold and there is no gauge freedom on that sector in the
supermembrane. In fact, there area preserving constraints generate
solely diffeomorphisms homotopic to the identity. In distinction
in the minimally immersed M2-brane sector that we
are considering there is an additional symmetry which may allow such
identification if desired.

The algebra of the first class constraint however does not close for
arbitrary values of $N$ and $M$. It closes for  $N,M$ finite or both
$N,M$ infinite, and those cases are considered below. The most
interesting case corresponding to $N$ infinite (i.e recovering the
continuum) with an arbitrary number of colors $M$ does not have a
closed algebra in this first construction. The modification needed
to hold is done in detail in the next Section.

\subsection{Some interesting limits}
We then have a $SU(N)\times SU(M)$ gauge model in $(0+1)$
dimensions, describing the correct number of degrees of freedom. We
may now consider different large $N$,$M$ limits to describe the
continuum.
\begin{itemize}
\item{The first case} we consider is when $N=M$, $N\to \infty$, we
have \bea
lim_{N\to\infty}\frac{1}{4iN}\mathbf{F}_{\mathbb{A}\mathbb{B}}^{\mathbb{C}}=\widetilde{\mathbf{F}}_{\mathbb{A}\mathbb{B}}^{\mathbb{C}} \eea by
redefining the interacting terms to eliminate the factor $2iN$ we
obtain the algebraic structure of $APD\times APD$. The hamiltonian
becomes \bea H\equiv \int d\sigma^2 \sqrt{W}\mathcal{H}, \eea
\bea\begin{aligned}
\mathcal{H}=&\frac{1}{2}(\frac{P^{ma}}{\sqrt{W}})^2+\frac{1}{2}(\frac{\Pi^{ra}}{\sqrt{W}})^2
+\frac{1}{4}(\{X^{mb},X^{nc}\}\widetilde{d}_{ab}^c+X^{mb}X^{nc}g_{ab}^c)^2+
\frac{1}{2}(\mathcal{D}_r X^{na})^2+\\
\nonumber
&\frac{1}{4}(\mathcal{F}_{rs}^a)^2-\overline{\Psi}\Gamma_{-}\Gamma_{r}D_r
\Psi-
\overline{\Psi}^a\Gamma_{-}\Gamma_m[\{X^{mb},\Psi^c\}\widetilde{d}_{bc}^a-
X^{mb}\Psi^c g_{bc}^a]\end{aligned}\eea and the constraint
\bea\begin{aligned} &\mathcal{D}_r
(\frac{\Pi^{ra}}{\sqrt{W}})+\{X^{mb},\frac{P_{m}^c}{\sqrt{W}}\}\widetilde{d}_{bc}^a+
X^{mb}(\frac{P_m^c}{\sqrt{W}})g_{bc}^a \\
\nonumber
&-\{\overline{\Psi}^b\Gamma_{-},\Psi^c\}\widetilde{d}_{bc}^a
-\overline{\Psi^b}\Gamma_{-}\Psi^c g_{bc}^a =0\end{aligned}\eea it
is also a first class constraint. In the above expression \bea
\mathcal{D}_r
\bullet^a=D_r\bullet^a+\{A_r^b,\bullet^{c}\}\widetilde{d}_{bc}^a +
A_{r}^b \bullet^{c}g_{bc}^a. \eea It is a well-known fact the
connection between large $SU(N)$ groups and the group of
diffeomorphims preserving the area \cite{deWit:1989vb} so this case
emerge naturally when APD is imposed in the action \footnote{We
thank T. Ortin for suggestive questions about this fact.}. In order
to recover the continuum limit it is implicitly imposed periodicity
on the new two arising dimensions, this is always the case when one takes the large M
limit. The result we obtain can be interpreted as the description of
condensate of M2-branes minimally immersed along those compact
surface. Since the two APD can be thought as orthogonal directions,
one can naturally think that they span a 4D surface related to a
M5-brane compactified on a $S^1$ with some appropriate fluxes and forms that
characterize the precise sector that we are considering.
The M5-brane hamiltonian on a $S^1$ expressed in terms Nambu algebras was obtained in \cite {DeCastro:2003jc}. The
exhaustive analysis required to determine precisely this statement
lies outside the scope of the present paper. In terms of a BLG formulation
it was already found a low energy description of a condensate
of multiple M2-branes by extending the number of colors to infinite
\cite{Bandos:2008jv}. In their analysis the remaining symmetry is a
Diff3-volume diffeomorphisms in distinction with ours that
correspond to the product of two APD.
Previously, in \cite{Bandos:2008fr,Ho:2008bn,Ho:2008nn,Bonelli:2008kh} using the BLG approximation it was also be pointed out its
natural connection with M5-brane worldvolume action.

\item{The SU(2) color case with $N$ infinite}  is a particular case since
$d_{ab}^c=0$ so we loose the symplectic structure that characterizes the M2-brane.
The hamiltonian becomes, \bea
\begin{aligned}
\mathcal{H}^{SU(2)}=&\frac{1}{2}(\frac{P^{ma}}{\sqrt{W}})^2+\frac{1}{2}(\frac{\Pi^{ra}}{\sqrt{W}})^2
+\frac{1}{4}(X^{mb}X^{nc}\epsilon_{ab}^c)+
\frac{1}{2}(\mathcal{D}_r X^{na})^2+\\
\nonumber
&\frac{1}{4}(\mathcal{F}^{SU(2)a}_{rs})^2-\overline{\Psi}\Gamma_{-}\Gamma_{r}D_r
\Psi- \overline{\Psi}^a\Gamma_{-}\Gamma_m X^{mb}\Psi^c
\epsilon_{bc}^a]\end{aligned}\ \eea and the constraint \bea\begin{aligned}
&\mathcal{D}_r (\frac{\Pi^{ra}}{\sqrt{W}})+
X^{mb}(\frac{P_m^c}{\sqrt{W}})\epsilon_{bc}^a
-\overline{\Psi^b}\Gamma_{-}\Psi^c \epsilon_{bc}^a =0\end{aligned}\eea is
the $SU(2)$ first class constraint and\bea \mathcal{D}_r^{SU(2)}
\Pi^{ra}=D_r\Pi^{ra}+A_{r}^b \Pi^{rc}\epsilon_{bc}^a. \eea So the
nonabelian product exactly corresponds to a SU(2) SYM without
corrections. Note that the covariant derivative $D_r$ in distinction
with the ordinary case is not just a partial derivative but inherits
the information of the global symplectic bundle through
$D_r=2\p R_rl_r\epsilon^{ab}\partial_a\widehat{X}_r\partial_b$.
\item{The third  case} we consider is when $N$ and $M$ are finite
but $N\gg M$ with $N$ large to be a good approximation
to the continuum limit although the case $N\to\infty$ does not satisfy the algebra.
The structure constant
\bea
\mathbf{F}_{\mathbb{A}\mathbb{B}}^{\mathbb{C}}=&NM\sin({\frac{(b\wedge
a)\pi}{M}+ \frac{(A\wedge B) \pi}{N}})
\eea
corresponds roughly speaking to the first two terms contributions of the complete expansion in Section 4,
however at this regularized level one cannot see the proper decoupling limit, so we leave this analysis to the next section.
One can characterize the qualitative properties of the quantum spectrum in this approximation.
Since the spectrum of the regularized hamiltonian of a single supermembrane minimally immersed MIM2
is purely discrete at bosonic and supersymmetric level, then it also holds for the $SU(N)\times SU(M)$ case as far as $N,M$ are both finite.\newline
 In the next section we would like to go further and analyze rigourously the case for N infinite
  with an arbitrary number of colors $M$ corresponding strictly to a stack of $M$ parallel MIM2-branes.
  \end{itemize}
\section{Multiple MIM2's with an arbitrary colors: a generalized star-product.}
In this section we obtain the $U(M)$ non-abelian extension of the MIM2 for an arbitrary number of colors $M$. We extend the algebraic symplectic structure of the
supermembrane with central charges in the L.C.G in terms of a
noncommutative product and a $U(M)$ gauge group. The main point is
to show that in such extension the original area preserving
constraint preserves the property of being first class.When an
abelian gauge group is considered the closure of the area preserving
constraints occurs with the complete noncommutative expansion as
well as  with the first two terms in the product expansion, the
exact symplectic structure \cite{Martin:2001zv}. In distinction when a $U(M)$
gauge group is considered there is only one possibility, the
complete noncommutative expansion. It is not enough to have the
symplectic structure tensor $U(M)$ in order to close the algebra of
the first class constraint. The complete expansion related to a
noncommutative associative product is needed. It is interesting that
this argument does not exclude an algebraic extension in terms of a
non-associative noncommutative product, which we will discuss
elsewhere. The noncommutative product we may introduce is
constructed with the symplectic two form already defined on the base
manifold $\Sigma$: \bea \omega_{ab}=\sqrt{W}\epsilon_{ab}, \eea
where $\sqrt{W}=\frac{1}{2}Area_{T^2}(\epsilon_{rs}\epsilon^{ab}\partial_a
\widehat{X}^r\partial_b \widehat{X}^s)$.  In this section,
in order to get a better insight on the star product, we use coordinates on the base manifold with length dimension $+1$
and define the dimensionless $\sqrt{W}$ with the area factor. All results of section 2 are of course valid.
 The two-form $\omega$ define
 the area element which is preserved by the diffeomorphisms
 generated by the first class constraint of the supermembrane theory
 in the Light Cone Gauge, which are homotopic to the
 identity, and by the $SL(2,Z)$ group of large diffeomorphisms discussed in section
 2.
 The two-form is closed and nondegenerate over $\Sigma$. By Darboux
 theorem one can choose coordinates on an open set $\mathfrak{N}$ in
 $\Sigma$  in a way that $\sqrt{W}$ becomes constant on $\mathfrak{N}$.
However this property cannot be extended to the whole compact
manifold $\Sigma$. The noncommutative theory must be globally constructed
from a non-constant symplectic $\omega$. The construction of such
noncommutative theories, for symplectic manifolds
was performed in \cite{Fedosov:1996fu,Fedosov:1994zz}. The general construction for Poisson
manifolds was obtained in \cite{Kontsevich:1997vb}. The Fedosov approach was
used to construct noncommutative Yang Mills Theories and also
noncommutative abelian membrane theories in \cite{Martin:2001zv}. A lot of work
on noncommutative Yang-Mills theories was developed for constant
$\omega$, some of them are \cite{Jurco:2001my,Jurco:2001rq}. See for example,
\cite{Wohlgenannt:2003de} for an
introductory review. We emphasize that our construction is not related to a Seiberg-Witten limit
of String Theory \cite{Seiberg:1999vs} in which one obtains a noncommutative theory with constant $B$-field.

The starting point on the Fedosov construction is a symplectic
manifold $(\Sigma,\omega)$ were $\omega$ is a symplectic two-form,
defining a symplectic structure on each tangent space
$T_{\sigma}\Sigma$. The elements of the Weyl algebra are formal
series \bea g(\xi,h)=\sum h^k g_{k,\alpha}\xi^\alpha \eea where $h$
is a parameter, $\xi\in T_{\sigma}M$ an $\alpha$ is a multi-index.
The associated product is defined by \bea g\circ f=\sum_{k=0}^\infty
(\frac{-ih}{2})^k\frac{1}{k!}\omega^{a_1 b_1}\dots \omega^{a_k
b_k}\frac{\partial^k
g}{\partial\xi^{a_1}\dots\partial\xi^{a_k}}\frac{\partial^k
f}{\partial\xi^{b_1}\dots\partial\xi^{b_k}}\eea where the terms are
ordered according to the weights $deg(\xi)=1,deg (h)=2$. In our construction
the $h$ parameter will be identified with the area wrapped by the membrane on the torus. The $i$
 factors are exactly the correct ones to reproduce the symplectic bracket with the same coefficients as in the abelian MIM2.
 The symplectic bracket will appear as the second term in the noncommutative bracket. The Weyl
algebra bundle $W$ is the union of the algebras $W_\sigma$,
$\sigma\in \Sigma$. Its sections are denoted $g=g(\sigma,\xi,h)$
where the coefficients $g_{k,\alpha}(\sigma)$ in the above expansions
are now covariant symmetric tensor fields on $\Sigma$. Differential
q-forms are naturally defined as a section of the bundle
$W\otimes\Lambda^q$. They constitute an algebra denoted
$C^{\infty}(W\otimes \Lambda)$. The commutator of two forms $g\in
W\otimes\Lambda^{q_1}$ and $f\in W\otimes\Lambda^{q_2}$ is defined
as \bea [g,f]_{\circ}=g\circ f-(-1)^{q_1q_2}f\circ g. \eea On any
symplectic manifold $\Sigma$ there always exist a torsion free
connection preserving the tensor $\omega$. The corresponding
covariant derivative will be denoted $D_a$, it satisfies $D_a
\omega_{bc}=0.$ Two symplectic connections differ by a completely
symmetric tensor. Given a symplectic connection on $\Sigma$, a
connection on $W\otimes\Lambda$ may be defined as \bea
Dg=d\sigma^b\wedge D_b g
 \eea
More general connections $\mathcal{D}$ are defined as \bea
\mathcal{D}g=Dg+\frac{i}{h}[\mathcal{A}, g]_{\circ} \eea where $\mathcal{A}$
is a section of $W\otimes \Lambda^1$. The next step in the Fedosov
construction is to introduce an Abelian connection: $\mathcal{D}$ is
Abelian if its curvature is a central form of the algebra, that is
\bea [\Omega,a]_{\circ}=0 \eea for any $a\in C^{\infty}(W\otimes \Lambda)$,
$\Omega$ is the curvature of $\mathcal{D}$. There always exist an
Abelian connection \cite{Fedosov:1994zz} in the Weyl algebra bundle. The Abelian
connection depends explicitly on the Riemann tensor of the
symplectic connection. The subalgebra $W_{abelian}\subset C^{\infty}
(W)$ of flat sections, that is the set of $g\in C^{\infty}(W)$ such
that $\mathcal{D}g=0$, where $\mathcal{D}$ is abelian is called the quantum algebra.

The center $Z$ of $W$ are the elements which do not depend on $\xi$.
For each section $g(\sigma,\xi,h)\in C^{\infty}(W)$, $\sigma(g)$
denotes the projection onto the center: \bea
\sigma(g(\sigma,\xi,h))=g(\sigma,0,h). \eea It follows that the
map $\sigma: W_{abelian}\to Z$ is bijective. Consequently a star
product $*$ on $C^{\infty}(\Sigma)$ may be defined as \bea
\widehat{g}*\widehat{f}=\sigma(\sigma^{-1}(\widehat{g})\circ
\sigma^{-1}(\widehat{f})) \eea for any $\widehat{g},\widehat{f}\in
Z$. The noncommutative star product is as a result of the
construction associative. The star product, for the particular case
in which $\omega$ has constant coefficients and the symplectic
connection is trivial, reduces to the Moyal product. In general
 the star product includes terms depending on the Riemann tensor
 for the symplectic connection. In particular for the symplectic
 structure on the base manifold of the supermembrane with central
 charges, the symplectic connection is necessarily non-trivial. For
 an explicit construction see \cite{Martin:2001zv}.
 We now extend the above construction and consider the tensor
 product of the Weyl algebra bundle times the enveloping algebra of
 $u(M)$. It may be constructed in terms of the Weyl-algebra
 generators $T_A$ introduced in the previous section, with the
 inclusion of the identity associated to $A=(0,0)$. This complete
 set of generators determine an associative algebra under matrix
 multiplication. The inclusion of the identity allows to realize
 the generators of the $u(M)$ in terms of $T_A$
 matrices, with $A=(a_1,a_2)$ and $a_1,a_2=-(M-1),\dots,0\dots M-1$. All the
 properties of the Fedosov construction remain valid, in particular
 the associativity of the star product. It is also valid the
 following Trace property, if $g=g^A T_A$, $f=f^AT_A$, $g^A,f^A\in C^{\infty}(W)$
 \bea\begin{aligned}
&Tr \int_{\Sigma }\sqrt{W}\sigma(g\circ
f)=\int_{\Sigma}\sqrt{W}\sigma(g^A\circ f^B)Tr(T_A
T_B)=Tr\int_{\Sigma}\sqrt{W}\sigma(f\circ g)\\ \nonumber &Tr
\int_{\Sigma} \sqrt{W}\sigma(g\circ f \circ h)= Tr \int_{\Sigma}
\sqrt{W}\sigma(h\circ g\circ
  f).\end{aligned}
 \eea
We may introduce canonical variables on the Weyl algebra bundle
\bea\begin{aligned}
&[X^m(\sigma,\xi,h),P_n\sigma^{'},\xi^{'},h)]_{P.B}=\delta_n^m\delta(\sigma^{'}-\sigma)\delta(\xi^{'}-\xi)\\
\nonumber & [A_{r}(\sigma,\xi,
h),\Pi^{s}(\sigma^{'},\xi^{'},h)]_{P.B}=\delta_r^s
\delta(\sigma^{'}-\sigma)\delta(\xi^{'}-\xi)\end{aligned} \eea It
then follows \bea &Tr\int_{\Sigma}\sqrt{W}\sigma(G\circ H\circ
[X^n),Tr\int_{\Sigma}\sigma (P_m]_{P.B}\circ L\circ
M)\\
\nonumber &=Tr\int_{\Sigma}\sqrt{W}\sigma(G\circ H\circ L\circ
M)\delta_m^n, \eea where we have used $Tr(T_C T_A T_B)Tr(T_D T_E
T_F)\eta^{CD}=Tr(T_AT_B T_E T_F)$ and the associativity of
the Weyl product.
In order to construct the hamiltonian of the theory we consider the
following connection on the Weyl bundle \cite{Martin:2001zv} \bea
\mathcal{D}\diamond=\frac{i}{h}[G_r
e^r,\diamond]_{\circ}+\frac{i}{h}[\mathcal{A}_r e^r,
\diamond]_{\circ}\eea where $G_r, \mathcal{A}_r\in C^{\infty}
(W_{Abelian})$, $\sigma G_r= \delta_{rs}X_h^s$ and $X_h^s= 2 \pi R_s
l_s \widehat{X}^s$. It corresponds to the harmonic sector of the map
to the compact sector of the target space. $\sigma
\mathcal{A}_r=A_r$ using the notation of section 2, $e^r=\partial_a
\widehat{X}^r d\sigma^a$. Its curvature is given by \bea\label{u2}
\Omega=\frac{i}{2h}[G,G]_{\circ}+\frac{i}{h}[G,\gamma]_{\circ}
+\frac{i}{2h}[\gamma,\gamma]_{\circ},\quad
\gamma=\mathcal{A}_r^Be^rT_B \eea

We now consider $(X^m,P_m), (A_r,\Pi^r)$ the canonical conjugate
pairs as well as the spinor fields $\Psi$ lifted to the quantum
algebra $W_{abelian}\in C^{\infty}(W)$. In order to simplify the
notation we use the same symbols for the lifted quantities. In the
presence of a $\circ$ product we refer to the lifted quantities. The
constraint is then defined as \bea\begin{aligned}
\phi(\sigma,\xi,h)\equiv &\mathcal{D}_r\frac{\Pi^{r}}{\sqrt{W}}+
\frac{i}{h}[X^{m},\frac{P_m}{\sqrt{W}}]_{\circ}+\frac{i}{h}[\overline{\Psi}\Gamma_{-},\Psi]_{\circ}\\
\nonumber & =\mathcal{D}_r \frac{\Pi^{rA}}{\sqrt{W}}T_A+
\frac{i}{h}(X^{mB}\circ
\frac{P_m^C}{\sqrt{W}}-\frac{P_m^B}{\sqrt{W}}\circ X^{C}_m)T_B T_C
+\frac{i}{h}[\overline{\Psi}\Gamma_{-},\Psi]_{\circ}.
\end{aligned}\eea with \bea\begin{aligned}
\mathcal{D}_r\frac{\Pi^{r}}{\sqrt{W}}&=\frac{i}{h}[G_r,\frac{\Pi^{r}}{\sqrt{W}}]_{\circ}+\frac{i}{h}[A_r,\frac{\Pi^r}{\sqrt{W}}]_{\circ}\\
\nonumber & = \frac{i}{h}[G_r,\frac{\Pi^{rA}}{\sqrt{W}}]_{\circ}T_A+ \frac{i}{h}(A^{B}_r\circ \frac{\Pi^{rC}}{\sqrt{W}}-\frac{\Pi^{rB}}{\sqrt{W}}\circ
A^{C}_r)T_B T_C.\end{aligned} \eea
 We notice that the first two terms of the commutator
\bea [X^{m},\frac{P_m}{\sqrt{W}}]_{\circ}=X^{mB}\frac{P_{m}^C}{\sqrt{W}} f_{BC}^E T_E
+(-i \frac{h}{2})\{ X^{mB},\frac{P_m^C}{\sqrt{W}}\}d_{BC}^E T_E+ O((h\omega)^2) \eea are
the terms which we considered in the previous section as extensions of the algebraic
structure of the supermembrane in the Light Cone Gauge. The
additional terms arising from the noncommutative product, ensuring
an associative product, are relevant in order to close
the constraint algebra. In fact using the trace properties discussed
above it follows \bea
&[Tr\int\sqrt{W}\sigma(\lambda(\sigma,\xi,h)\circ\phi(\sigma,\xi,h)),
Tr\int\sqrt{W}\sigma(\epsilon(\sigma,\xi,h)\circ
\phi(\sigma,\xi,h))]_{P.B}\\ \nonumber &
 =Tr\int\sqrt{W}\sigma([\lambda,\epsilon]_{\circ}\circ\phi(\sigma,\xi.h)),\eea
$\phi\in W_{abelian}$ is a first class constraint generating a gauge
transformation which is a deformation of the original are preserving
diffeomorphisms. In particular, \bea
\sigma[-i\int\sqrt{W}(\lambda\circ
\phi),X^m]_{P.B}=\frac{i}{h}\sigma[\lambda,X^m]_{\circ}=\frac{1}{2}\{\lambda,
X^m\}+O((h\omega)^2), \eea where $\lambda$ is the infinitesimal parameter of the area preserving diffeomorphisms,
valued on the generator $T_{(0,0)}$. \newline


The projection of $\Omega$ in (\ref{u2}) has the expression
\cite{Martin:2001zv} \bea \sigma\Omega=-\omega +\mathcal{F}+ O(h^2)
\eea where \bea \mathcal{F}=\frac{1}{2}e^r\wedge e^s (D_r A_s -D_s
A_r +\frac{i}h{\{A_r,A_s\}}_{\ast}), \eea and
$\omega=\frac{1}{2h}\sqrt{w}\epsilon_{ab}d\sigma^a \wedge
d\sigma^b$. $O(h^2)$ depend explicitly on the Riemann tensor of the
symplectic connection. $D_r,D_s$ are the ones defined in section 2.
The hamiltonian of the theory for  $M$ multiple parallel M2-branes
with $U(N)$ gauge group is then \bea\begin{aligned}
Tr\int_{\Sigma}\mathcal{H}=Tr\int_{\Sigma}\sqrt{W}&\Bigg[\frac{1}{2}(\frac{P^m}{\sqrt{W}})^2
+\frac{1}{2}(\frac{\Pi^r}{\sqrt{W}})^2 +\frac{1}{2h^2}(\{X_h^r,
X^m\}_{\ast}+\{A^r,X^m\}_{\ast})^2\\ \nonumber &
+\frac{1}{4h^2}\{X^m,X^n\}^2_{\ast}
+\frac{1}{2}\Omega_{rs}\Omega^{rs}\\
\nonumber & - \frac{i}{h}\overline{\Psi}\Gamma_{-}\Gamma_{r}
(\{X_h^r, \Psi\}_{\ast}+\{A^r,\Psi\}_{\ast})
-\frac{i}{h}\overline{\Psi}\Gamma_{-}
\Gamma_{m}\{X^{m},\Psi\}_{\ast}\Bigg],\end{aligned}\eea where the
term
$\{{X}^r_h,X^m\}_{\ast}+\{A^r,X^m\}_{\ast}=\delta^{rs}\mathcal{D}_s
X^m +O(h)$ in the notation of section 2. The hamiltonian is subject
to the first class constraint \bea \phi\equiv \{X_h^r,
\frac{\Pi^r}{\sqrt{W}}\}_{\ast}+\{ A_r,
\frac{\Pi^r}{\sqrt{W}}\}_{\ast}
+\{X^m,\frac{P_m}{\sqrt{W}}\}_{\ast}-\{\overline{\Psi}\Gamma_{-},\Psi\}_{\ast}=0
\eea The first terms in the star product expansion are \bea
\phi\equiv \mathcal{D}_r\frac{\Pi^r}{\sqrt{W}}+\{X^m,
\frac{P_m}{\sqrt{W}}\}-\{\overline{\Psi}\Gamma_{-},\Psi\}+ O(h) \eea
where $\{,\}_{\ast}$ has been normalized in a way to be a
deformation of $\{,\}$ the symplectic bracket of the supermembrane
in the L.C.G. The fields are now $u(M)$ valued. The constraint is a
deformation of the $u(M)$ Yang-Mills constraint. The terms $O(h)$
involve the Riemann tensor of the symplectic connection which itself
depends on the symplectic two-form introduced by the central charge.
The $O(h)$ terms are necessary in order to close the constraint
algebra. An explicit expression for the curvature in the abelian
case was found in \cite{Martin:2001zv},

\bea\begin{aligned}
\sigma\Omega=&-\omega+\mathcal{F}-\frac{h^2}{96}
(R_{bcda}\big(D_{\widehat{b}}D_{\widehat{c}}D_{\widehat{d}})A_m
-\frac{1}{4}R_{\widehat{b}\widehat{c}\widehat{d}p}\epsilon^{pq}D_q
A_m\big)\epsilon^{b\widehat{b}}\epsilon^{c\widehat{c}}\epsilon^{d\widehat{d}}e^a\wedge
e^m \\ \nonumber
&-\frac{h^2}{96.8}R_{bcda}R_{\widehat{b}\widehat{c}\widehat{d}m}
\epsilon^{b\widehat{b}}\epsilon^{c\widehat{c}}\epsilon^{d\widehat{d}}e^a\wedge
e^m+O(h^3)\dots.\end{aligned}\eea

The terms involving the Riemann
tensor of the symplectic connection are absent in the Moyal product.

If we make manifest the dimensional dependence of the star-product we can realize
 that the parameter $[h]=n. Area_{T^2}$, $n$ is the wrapping number.
 In fact, as it was introduced in the definition of the noncommutative product, $h$ has degree $2$ while $\xi$ has degree $1$. Since $\xi$ has legth dimension 1 then $h$ must have length dimension $2$ in order to have a dimensionless noncommutative product.

The star-product  is explicitly given by
\bea
\begin{aligned}
\frac{i}{h}\{f,g \}^a_{\ast}= & \frac{i}{h}f^b g^c f_{bc}^a + \{f^b,g^c\}d_{bc}^a+ O(h)
\\ \nonumber &= \frac{i}{n Area_{T^2}}f^b g^c f_{bc}^a+\{f^b,g^c\}d_{bc}^a+ O(n Area_{T^2})
\end{aligned}
\eea

where $\{f^b,g^c\}=\epsilon^{rs}D_r f^b D_s g^c$. $D_r$ was defined in section 2. The factor $\frac{1}{h}$ ensures that this formalism is a
 nonabelian extension of the abelian MIM2-brane,  since for the abelian case
$f_{bc}^a$ vanishes, $d_{bc}^a=1$, and the algebra closes exactly with the ordinary symplectic
bracket corresponding to a single M2 action without further contributions.

The mass square operator may be written as: \bea -mass^2=\int
(\frac{1}{2}d\widehat{x}^r\wedge
d\widehat{x}^s\epsilon_{rs})Tr[\frac{1}{2}(\frac{P}{\sqrt{W}})^2
+\frac{1}{2}(\frac{\Pi}{\sqrt{W}})^2+(T Area_{T^2}^2)(V_B+V_F))]
\eea where $V_B$ and $ V_F$ are the bosonic and fermionic potentials
of the Hamiltonian. The scale of the theory is then
$T.n.\textrm{Area}_T^2$. The measure of integration reduces to the
dimensionless \newline $\frac{1}{2}d\widehat{x}^r\wedge
d\widehat{x}^s\epsilon_{rs}$. The conjugate momenta have mass
dimension $+1$, and the corresponding configuration variable mass
dimension $-1$. $T$ has mass dimension $+3$. On the other hand,
 by considering the contribution to Yang-Mills arising from the first term in the
above expansion of the star product and by taking canonical
dimensions for the conjugate pairs we get for the coupling constant

\bea
g_{YM}=\frac{1}{T_{M2}^{1/2}. n. Area_{T^2}}.
\eea
It has dimension of $mass^{1/2}$.


It represents the
coupling constant of the first term in the star-product expansion.
We assume that the compactification radii  is $R_i>>l_p$
but with the theory still defined at high energies.
For a fixed tension and winding number $n$,
the only relevant contribution in the star product  at low energies is the $U(M)$ commutator
since the natural length is much larger larger that the effective radii $R_{eff}=n^{1/2}\sqrt{R_1R_2}$.
This is the decoupling limit of the theory since the Yang Mills field strength becomes the coupling constant of the theory and gravitational modes become decoupled. The $g_{YM}$ is very large in this phase and the theory is in the IR phase.
It corresponds to have a description of  M multiple MIM2-branes as point-like particles, representing $M$ the number of supermembranes.
As we raise the energy the $g_{YM}$ coupling constant gets
weaker and for energies high enough, comparable with the natural scale of a MIM2-brane with an effective area of ($n.Area_{T^2}$),
the oscillation and vibrational modes containing the gauge but also gravity interactions between the supermembranes are no longer negligible so the full star- expansion has to be considered.
All terms associated to the supermembrane symplectic structure of the star-bracket contribute while the ordinary SYM contribution vanishes.
The point-like particle picture is no longer valid, and it is substituted for that of an extended $(2+1)$D object and the gauge and gravity contributions are strongly coupled.
One can define formally and effective physical coupling constant for the ordinary $F_{\mu\nu}$ field strength which it would correspond to
$\Lambda=M. g_{YM}$ with $M$ representing the number of supermembranes and then one can try to obtain the 't Hooft coupling expansion in the large M.
In this picture however one should take care on the limit.
By keeping $\Lambda$  fixed with $M$ going to infinity, for a fixed tension and a fixed compactification radii, one has to consider the wrapping number $n$ also going to infinity.
But $n.Area$ is the order parameter that would also go multiplied by $M$ in the expansion so one enters "faster" in the strong correlated
limit where the rest of the terms of the star-product expansion cannot be neglected, moreover, from a physical point of view
$n. area_{T^2}$ is related to the size of the MIM2 as an extended object and it cannot be larger than the present energy bounds we have, otherwise
it would be in contradiction with our point-particle description at low scales.
In order to perform a more accurate analysis one should be working
with the nonabelian extension of the MIM2 for 4D noncompact,-it will be considered elsewhere- however we believe that the qualitative arguments
presented here should remain valid also in that case.
\section{ $N=8$  LCG Supersymmetry}
 In order to analyze the invariance of the MIM2 action under supersymmetry it is convenient to introduce,
\bea
\begin{aligned}
&\mathcal{D}_r \cdot = [\widetilde{X}_r, \cdot ]_{\ast}+ [A_r, \cdot]_{\ast}\\ \nonumber &
\mathcal{D}_m \cdot= [X_m,\cdot]_{\ast}\\ \nonumber &
\mathcal{D}_0 \cdot= \partial_{\tau} \cdot +[A_0, \cdot]_{\ast}
\end{aligned}
\eea
where $A_0$ is the lagrange multiplier associated to the first
class constraint and $\widetilde{X}_r=X_h^s\delta_{rs}$ with $r=1,2$. We will denote $A_m=X_m$.
We then introduce the index $\mu=0,r,m$.  The $\mathcal{D}_{\mu}$ satisfy the Leibniz rules
\bea
\begin{aligned}
&\mathcal{D}_{\mu}F\circ G=\mathcal{D}_{\mu}F\circ G
+F\circ \mathcal{D}_{\mu}G \\ \nonumber &
\mathcal{D}_{\mu}[F,G]_{\circ}= [\mathcal{D}_{\mu}F, G]_{\circ}+[F,\mathcal{D}_{\mu} G]_{\circ}
\end{aligned}
\eea
We also consider the curvatures
\bea \label{es1}
\begin{aligned}
&\Omega_{rs}=[\widetilde{X}_r, A_s]_{\ast}-[\widetilde{X}_{s},A_r]_{\ast} +[A_r,A_s]_{\ast}+ [\widetilde{X}_r,\widetilde{X}_s]_{\ast}\\ \nonumber &
\Omega_{rm}=\mathcal{D}_r X_m \\ \nonumber &
\Omega_{mn}=[X_m,X_n]_{\ast}\\ \nonumber &
\Omega_{0r}=\dot{A}_r-[\widetilde{X}_r,A_0]_{\ast}+[A_0, A_r]_{\ast}\\ \nonumber &
\Omega_{0m}=\mathcal{D}_0 X_m.
\end{aligned}
\eea

$\Omega_{\mu\nu}$ satisfy the Bianchi identities
\bea
\mathcal{D}_{\mu}\Omega_{\nu\lambda}+\mathcal{D}_{\lambda}\Omega_{\mu\nu}
+\mathcal{D}_{\nu}\Omega_{\lambda\mu}=0
\eea
These relations are valid provided for any associative product.
We are considering $A_r,X_m$ valued in the enveloping algebra of $U(N)$ in terms of the Weyl-Heisenberg
generators $T_A$ of section 3, $\circ$ is the noncommutative associative Weyl product. The  tensor
product is still an associative product and the property (\ref{es1}) is satisfied identically in our construction.
The lagrangian of the theory after integration of the momenta $P_m, \Pi^r$ maybe expressed as
\bea
\mathcal{L}=-\frac{1}{4}\Omega_{\mu\nu}\ast \Omega^{\mu\nu}
-\overline{\Psi}\ast \Gamma_{-}\Gamma^{M}\mathcal{D}_M \Psi
- \overline{\Psi}\ast\Gamma_{-}\mathcal{D}_0 \Psi\eea
where $M=r,m$. The light cone fermionic gauge condition we use is
\bea
\Gamma_+ \Psi=0.
\eea
The associated action is invariant under the following
supersymmetric transformations with parameter $\epsilon=\Gamma_-\Gamma_+ \epsilon$
\bea
\begin{aligned}
&\delta A_M=\delta A_M^B T_B=\overline{\epsilon }\Gamma_M \Psi^B T_B \quad M=r,m\\ \nonumber &
\delta A_0 =\delta A_0^B T_B= -\overline{\epsilon}\Psi^B T_B \\ \nonumber &
\delta \Psi= \delta \Psi^B T_B= \frac{1}{4}\Gamma_{+}\Omega_{MN}^B \Gamma^{MN}\epsilon T_B
+ \frac{1}{2}\Gamma_{+}\Omega_{0M}^{B}\Gamma^M \epsilon T_B
\end{aligned}
\eea
These transformations are a $U(N)$ extension of the SUSY transformations for the supermembrane in the LCG found in
\cite{deWit:1988ig, Bergshoeff:1987qx} and they realize $\mathcal{N}=8$ supersymmetries on the worldvolume.
The invariance of the action arises in a similar way as it does for Super Yang-Mills:
\bea
\begin{aligned}
Tr\langle\delta(-\frac{1}{4}\Omega_{\mu}\ast \Omega^{\mu\nu})\rangle
&=Tr\langle \delta A_{\nu}\ast \mathcal{D}_{\mu}\Omega^{\mu\nu}\rangle \\\nonumber &
=Tr\langle 2\overline{\Psi}\ast\Gamma_{-}\Gamma^M\mathcal{D}_{M}\delta\Psi
+2\overline{\Psi}\ast\Gamma_{-}\mathcal{D}_0 \delta\Psi\rangle
\end{aligned}
\eea
and
\bea Tr\langle\overline{\Psi}\ast \Gamma_-\Gamma^M (\delta \mathcal{D}_M)\Psi\rangle=0,\eea

where  $\langle\cdot\rangle$ denotes integration and
 Bianchi identities as well as eleven dimensional identities for the $\Gamma$ matrix have been used.

\section{Discussion and Conclusions}
\subsection{Main results}
We have obtained a $N=8$ nonabelian  U(M) formulation of the
 minimally immersed supermembrane for arbitrary number of colors $M$ with
 all the symmetries of the supermembrane, in the LCG.
This corresponds to the M-theory
dual of the nonabelian Dirac-Born-Infeld theory,
representing a  bundle of D2-D0 branes. It is the first time that
a nonabelian gauge theory can be directly obtained from a full-fledged  sector of M-theory
element, so far restricted to String theory: Heterotics and
 Dp-branes in type II theories. This opens a new interesting
window for models in phenomenology.
At energies of the order of the compactification scale, the theory has the gauge and gravity sector strongly coupled.
It describes all of the oscillations modes of the multiple parallel M2-branes minimally immersed.
At low energies the theory enters in a decoupling regime and the physics is then described
 by a $\mathcal{N}=8$ SYM theory of point-like particles in the IR phase. We then expect to describe correctly many aspects of
phenomenology when realistic gauge groups are considered.  From the point of view of the target space the
theory has N=1 susy in 9D flat-dimensions. In \cite{GarciadelMoral:2007xj}
a N=1 target space, D=4 formulation of a single
supermembrane minimally immersed together with a number of
interesting phenomenological properties were found. Moreover in \cite{Belhaj:2008qz} a
formulation of the supermembrane minimally immersed on a G2 manifold
was also obtained. Its quantum supersymmetric spectrum is also
purely discrete. The analysis in 4D can be also extended to the
nonabelian case following the lines shown in this paper,
allowing to obtain models with reduced number of target and worldvolume supersymmetries.
\subsection{The case of M2 without central charge condition}
We could expect to apply the mechanism we have implemented to the supermembrane
without central charges on flat space $M_{11}$. However this is not possible. In fact the
nonabelian formulation of multiple parallel MIM2 has a constraint
 (\ref{em1}) of the  Gauss type allowing the elimination of the gauge degrees of freedom a la Yang-Mills. This constraint also generates, when the parameter
 is valued on the $T_{(0,0)}$ generator, a deformation of an area preserving diffeomorphims. For a single MIM2 the $U(1)$
  Gauss constraint and area preserving
  diffeomorphism constraint are the same,
\bea\label{em1}
\phi\equiv \{\widehat{X}_r, \frac{\Pi^r}{\sqrt{W}}\}_{\ast}+\{ A_r, \frac{\Pi^r}{\sqrt{W}}\}_{\ast}
+\{X^m,\frac{\Pi^r}{\sqrt{W}}\}_{\ast}-\{\overline{\Psi}\Gamma_{-},\Psi\}_{\ast}=0.
\eea
This does not happen for the 11D
supermembrane where the constraint is of the form \bea\label{e3b} \{X^m,P_m\}_{\ast}+ fermions=0.\eea
In (\ref{em1}), where a linear term is present, it is possible to eliminate exactly one degree of
freedom (with internal index) but in (\ref{e3b}) the
 gauge degrees of freedom cannot be eliminated correctly, unless an additional assumption is made.
 It is
not possible to impose the static gauge for the supermembrane on compact base manifold \cite{GarciadelMoral:2008qe}. It would imply
automatically the vanishing of string-like spikes that are present
in the formulation, and the spectrum of the hamiltonian would not be continuous.
\subsection{General properties of our construction}
 \begin{itemize} \item{} Our action starts from the
formulation of the supermembrane with a topological condition and
extends it to include nonabelian  $U(N)$ interactions while preserving all of
the symmetries and constraints of the theory. This means that we work at high energies
where gauge fields and supergravity contribution are strongly correlated. The analysis is then valid for any number of M2-branes large or small.

\item{} We are mainly interested in characterizing the M2-branes interactions by themselves.
The theory is not scale invariant so there is no a
sextic scalar potential in the action,  it remains quartic in the
fields although now, valued on a $U(N)$ algebra.

\item{The single MIM2 action} contains a gauge field invariant under
symplectomorphisms defined on the (2+1)D worldvolume action of the
supermembrane. The first class constraint generating the
symplectomorphisms is analogous to the Gauss constraint of SYM theory.
It is this property which allows the $U(N)$ formulation without changing
the number of degrees of freedom.
In the low energy description of the 11D
supermembrane the situation is different since it only contains scalar fields in the
bosonic sector.  For that case, in order to introduce a gauge field  without changing the degrees of freedom a
 Chern-Simons type extension is needed.

\item{} We are describing a supersymmetric theory of the multiple M2-branes (2+1)D
(with gauge and gravity sector coupled) strongly
coupled immersed in a $M_9$x$T^2$ target space with N=1 supersymmetry.
Quantum corrections are relevant in this case and its spectral properties are controlled by the bosonic sector.
The abelian MIM2 has a
well-defined supersymmetric quantum formulation in terms of a Feynman path integral.
We have argued that this property remains valid in the
nonabelian regularized version with the ordinary product.

 \item{}We may perform a matrix regularization of the proposed
  action of multiple M2-branes minimally immersed in the compactified space. We are
 able to do it since the central charge condition allow a proper
 treatment of the harmonic forms present on actions formulated on compactified spaces \cite{GarciadelMoral:2001zb}.

 \item{Our} approach is not covariant. It is a LCG formulation. However, remarkably,
  the bosonic sector of the theory corresponds to a non-standard reduction of
  SYM  with the star- product defined in Section 4, from 10D to $(2+1)$D.
  This is so, because the light cone coordinates decouple in our sector of the theory, since we are able to
 solve completely the constraint for $X^-$  once the gauge fixing condition has been imposed.
  The global constraint (\ref{e3}) guarantee that there is no winding condition on $X^{-}$.
  The LCG supersymmetric algebra has 8 generators, it is related to the one found in \cite{deWit:1988ig}
 for the 11D abelian supermembrane.
\end{itemize}
\subsection{Nonconstant star-product}
We would like to emphasize that
our approach is not related to a noncommutative Seiberg-Witten (SW) limit of String theory.
In order to close the algebra of  constrains while keeping the invariance of the action, we introduced a Fedosov
star-product. It is not related to the constant B-field of SW.
 In fact the SW formulation is a local description, by Darboux theorem of the Fedosov star-product.
Although our construction is based on a compactification on a
2-torus it does not
correspond to the theory on the noncommutative torus done
in \cite{Connes:1997cr}.
For the abelian formulation of the MIM2 we do not need to introduce an star-product
since the constraint closes at first order.
In the nonabelian case the constraint still closes at first order iff the number of colors go to infinite or iff
we restrict our analysis to a regularized case for an arbitrary number of colors. At exact level formulation (not regularized)
of the MIM2's with an arbitrary number of colors we need to
introduce a Fedosov the star-product to close the algebra.
This gives a precise indication on how U(N) gauge group is compatible with
area preserving diffeomorphisms. The particular case
of $SU(2)$ corresponds to the case where the symplectic structure vanishes and
it corresponds to a ordinary-type of SYM description.
It would be nice to see the connection, if any, between our results and the fact that
for low energy descriptions of M2-branes
only $N=8$ models has been only found for $SU(2)\times SU(2)$ \cite{Bagger:2007jr,VanRaamsdonk:2008ft}
gauge group or infinite gauge group representing a condensate of M2-branes\cite{Bandos:2008jv},
meanwhile the low energy formulation of a multiple M2 inspired action with an  arbitrary number of colors $U(N)\times U(N)$
has only been found for less number of supersymmetries $\mathcal{N}=6$ \cite{Aharony:2008ug}.

\section{Acknowledgements}
The authors would like to thank to I. Bandos, O. Bergman, M. Billo,
F. Cachazo, M. Frau, F. Gliozzi, A. Lerda, L. Magnea, I. Martin, H.
Nicolai, T. Ortin, D. Rodriguez-Gomez, J. Rosseel and M. Wolhgennant for
interesting discussions and comments.  MPGM is partially supported by the European
Comunity's Human Potential Programme under contract
MRTN-CT-2004-005104 and by the Italian MUR under contracts
PRIN-2005023102 and PRIN-2005024045 and by the Spanish Ministerio de
Ciencia e Innovaci\'on (FPA2006-09199) and the Consolider-Ingenio
2010 Programme CPAN (CSD2007-00042). The work of AR is partially  by
PROSUL, under contract CNPq 490134/2006-08.

\bibliographystyle{JHEP}

\bibliography{qk}

\end{document}